\documentclass[aps,prl,twocolumn,amsmath,amssymb,nofootinbib,showpacs]{revtex4}

\usepackage[english]{babel}
\usepackage{latexsym}
\usepackage{graphics}
\usepackage{epsfig}
\usepackage{color}
\usepackage{bm}
\usepackage{amsmath}
\usepackage{amssymb}

\begin{document}

\title{Three-Body Interacting Bosons in Free Space}

\author{D.~S.~Petrov}
\affiliation{Universit\'e Paris-Sud, CNRS, LPTMS, UMR8626, Orsay, F-91405, France}

\date{\today}

\begin{abstract}

We propose a method of controlling two- and three-body interactions in an ultracold Bose gas in any dimension. The method requires us to have two coupled internal single-particle states split in energy such that the upper state is occupied virtually but amply during collisions. By varying system parameters one can switch off the two-body interaction while maintaining a strong three-body one. The mechanism can be implemented for dipolar bosons in the bilayer configuration with tunneling or in an atomic system by using radio-frequency fields to couple two hyperfine states. One can then aim to observe a purely three-body interacting gas, dilute self-trapped droplets, the paired superfluid phase, Pfaffian state, and other exotic phenomena.

\end{abstract}


\maketitle

The Feshbach resonance technique, which allows for tuning the two-body interaction to any value, has been a major breakthrough in the field of quantum gases \cite{ChinRMP}. Reaching strongly interacting regimes by using this method is proven successful in two-component fermionic mixtures \cite{ZwergerBook} because of the naturally built-in mechanism of suppression of local three-body inelastic processes \cite{PSS}: the Pauli principle prohibits three fermions to be close to each other as at least two of them are identical. Essentially the same mechanism is responsible for the repulsion between weakly bound molecules in this system ensuring the mechanical stability. Bosons, having no such protection, are much more fragile. A Bose-Einstein condensate (BEC) collapses in free space even for infinitesimally weak attraction \cite{DalfovoRMP} not to mention devastating recombination losses close to resonance \cite{Inouye1998,Roberts2000}.

A repulsive three-body force can stabilize the system and induce nontrivial many-body effects. A weakly interacting BEC with two-body attraction (coupling constant $g_2<0$) and three-body repulsion ($g_3>0$) is predicted to be a droplet, the density of which in the absence of external confinement and neglecting the surface tension is flat and equals $n= -3g_2/2g_3$ \cite{Bulgac,RemLowDSolitons,Gammal}. For a strong (beyond mean-field) two-body attraction the spinless Bose gas can pass from the atomic to paired superfluid phase via an Ising-type transition with peculiar topological properties \cite{Radzihovsky,Stoof,Lee}. However, the mechanical stability of the system requires repulsive few-body interactions \cite{Mueller} or other stabilizing mechanisms \cite{Nozieres}. Few-body forces are also important for quantum Hall problems: the exact ground state of bosons in the lowest Landau level with a repulsive three-body contact interaction is the Pfaffian state \cite{Greiter} also known as the weak-pairing phase and characterized by non-Abelian excitations \cite{MooreRead}. Interestingly, a finite range of the three-body interaction breaks the pairing \cite{Jain}. On the other hand, there may be a transition from the weak- to strong-pairing Abelian phase \cite{ReadGreen}, presumably driven by varying $g_2$. 

Most proposals for generating effective three-body interactions deal with lattice systems \cite{Buchler,DaleyZoller,Mazza,Tiesinga,Taylor,Daley}. In free space, since three-body effects are significant when $g_3n$ is of order $g_2$, staying in the dilute regime requires small $g_2$ and large $g_3$. In three dimensions a resonant three-body force is predicted for large and negative scattering lengths when a three-body Efimov state crosses the three-atom threshold \cite{Efimov,Mehen,Bulgac,RemFeshbach}. This method is associated with strong relaxation losses \cite{Ferlaino}, although nonconservative three-body interactions can also lead to interesting effects \cite{DaleyZoller,Roncaglia}.

This Letter is motivated by the observation that dipolar particles trapped on a single layer and oriented perpendicular to the plane repel each other, whereas in a bilayer configuration there is always a bound state \cite{Simon,Shih,Armstrong,Klawunn,Baranov}. We argue that the bound state emerges from the scattering continuum as one gradually splits the layer into two and reduces the interlayer tunneling amplitude below a critical value $t=t_c$. We show that near this point $g_2\propto t-t_c$ providing the desired control over the two-body interaction. Next, we find that the three-body interaction near this zero crossing is repulsive and conservative: similar to the fermionic Pauli protection three dipoles are frustrated in the sense that at least two of them are on the same layer and, therefore, experience repulsion. Surprising and counterintuitive is that the effective three-body repulsion strengthens with decreasing the dipole-dipole interaction and interlayer tunneling amplitude. The reason is that the scattering wave function of two dipoles contains a significant contribution of a virtually excited interlayer dimer state of size $\sim 1/\sqrt{t}$. The effective three-body force originates from the interaction of the third particle with this state and becomes stronger with decreasing $t$. Based on this understanding we propose a general method of controlling few-body interactions applicable for atomic systems in any dimension.

The Hamiltonian of bosonic dipoles in the bilayer geometry with tunneling is written as (cf. \cite{Wang}) 
\begin{eqnarray}\label{Ham}
H&=& \int_{\pmb{r}}\sum_{\sigma} \Psi^\dagger_{\sigma {\boldsymbol{r}}} (-\nabla^2_{\boldsymbol{r}}/2+t)\Psi_{\sigma {\boldsymbol{r}}} -t (\Psi^\dagger_{\uparrow {\boldsymbol{r}}}\Psi_{\downarrow {\boldsymbol{r}}}+{\rm H.c.})\nonumber\\
&&\hspace{-1.cm}+\frac{1}{2} \int_{{\boldsymbol{r}},{\boldsymbol{r}}'}\sum_{\sigma,\sigma'} \Psi^\dagger_{\sigma {\boldsymbol{r}}} \Psi^\dagger_{\sigma' {\boldsymbol{r}}'}V_{\sigma\sigma'}(|{\boldsymbol{r}}-{\boldsymbol{r}}'|)\Psi_{\sigma {\boldsymbol{r}}} \Psi_{\sigma' {\boldsymbol{r}}'},
\end{eqnarray} 
where $\Psi^\dagger_{\sigma {\boldsymbol{r}}}$ is the creation operator of a boson on layer $\sigma(=\uparrow,\downarrow)$ with in-plane coordinate ${\boldsymbol{r}}$ and we neglect the transverse extension of the wave function within the layers compared to the interlayer distance $\lambda$. We adopt the units $\lambda=\hbar=m=1$. Then, for dipoles oriented perpendicular to the plane the intralayer and interlayer potentials equal $V_{\sigma\sigma}(r)=r_{*}/r^3$ and $V_{\sigma\sigma'}(r)|_{\sigma\neq\sigma'}=r_{*}(r^2-2)/(r^2+1)^{5/2}$, respectively, and $r_{*}$ is the characteristic length scale of the dipole-dipole potential.

\begin{figure}
\centerline{\includegraphics[width=0.9\hsize,clip,angle=0]{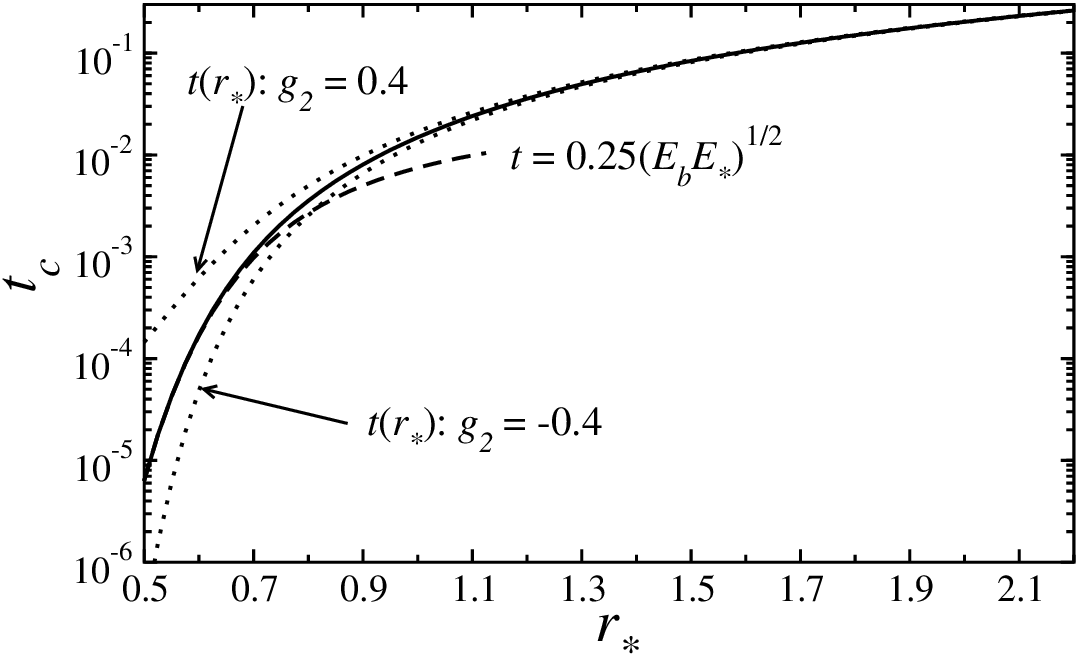}}
\caption{Critical tunneling amplitude $t_c$ vs. $r_*$ (solid line). For $t<t_c$ the two-body interaction is attractive and supports a bound state. The dotted lines enclose the region $-0.4<g_2<0.4$. The dashed line is the prediction of the zero-range theory $t_c\approx\sqrt{E_{\uparrow\downarrow}E_{\uparrow\uparrow}}/4$ valid for small $r_*$. }
\label{fig:tVSrs}
\end{figure}

The one-body spectrum of (\ref{Ham}) consists of two branches with dispersions $\varepsilon_+(k)=k^2/2$ and $\varepsilon_-(k)=2t+k^2/2$. The corresponding eigenfunctions are $\phi_{\pm,{\bf k}}=|\pm\rangle \exp(i{\bf k}\cdot {\boldsymbol{r}})$ with spinor parts $|\pm\rangle=(|\uparrow\rangle\pm|\downarrow\rangle)/\sqrt{2}$. We assume that the temperature and typical interaction scales are lower than $t$ so that the upper branch is excited only virtually during collisions. Moreover, due to the $\uparrow$-$\downarrow$ symmetry, the Hamiltonian (\ref{Ham}) couples the lowest two-body spinor configuration $(|+\rangle)^2$ only to the highest one, $(|-\rangle)^2$. The gap is then effectively $4t$. Writing the wave function of the relative motion in the form $(|\uparrow\rangle |\uparrow\rangle+|\downarrow\rangle |\downarrow\rangle)\phi_{\uparrow\uparrow}({\boldsymbol{r}})+(|\uparrow\rangle|\downarrow\rangle + |\downarrow\rangle|\uparrow\rangle)\phi_{\uparrow\downarrow}({\boldsymbol{r}})$ we obtain the two-channel Schr\"odinger equation
\begin{equation}\label{2channelSchr}
\left[-\nabla^2_{\boldsymbol{r}}-E+\left(\begin{array}{lr}V_{\uparrow\uparrow}(r)+2t& -2t\\
-2t&V_{\uparrow\downarrow}(r)+2t\end{array}\right)\right]\left(\begin{array}{lr}\phi_{\uparrow\uparrow}\\
\phi_{\uparrow\downarrow}\end{array}\right)=0.
\end{equation}
The bound state in the potential $V_{\uparrow\downarrow}$ \cite{Simon,Shih,Armstrong,Klawunn,Baranov,RemSingleBoundState} is a true eigenstate of our model in the limit $t\rightarrow 0$. Its wave function contains only the $\phi_{\uparrow\downarrow}$ part; i.e., the two dipoles are localized on different layers. The binding survives small $t$ since the cost of this localization is of order the (small) tunneling energy. However, for $t>t_c$ the localization becomes too expensive and the bound state crosses the two-particle threshold. Solid line in Fig.~\ref{fig:tVSrs} shows $t_c$ as a function of $r_*$ obtained numerically from Eq.~(\ref{2channelSchr}).


The low energy and small momentum properties of this rather unusual two-dimensional scattering problem with long-range interactions are described in terms of the vertex function $\Gamma(E,{\bf k},{\bf k}')$, where $E$ is the total energy in the center of mass reference frame and ${\bf k}$ and ${\bf k}'$ are the incoming and outgoing relative momenta, respectively. For sufficiently weakly bound or quasibound state, i.e., when $t$ is close to $t_c$, we can repeat arguments of Ref.~\cite{Baranov} and write
\begin{equation}\label{Vertex}
\Gamma(E,{\bf k},{\bf k}')\approx \frac{4\pi}{\ln(4t/E)+4\pi/g_2+i\pi}-2\pi r_*|{\bf k}-{\bf k}'|,
\end{equation}   
which, in our case, is valid for $E\sim k^2\sim k'^2\ll t$. The second term in Eq.~(\ref{Vertex}) accounts for the long-range dipole-dipole interaction tail and the first one is the usual single-pole expression for the scattering amplitude, which effectively integrates out the short-range radial motion and $\sigma$ degrees of freedom of Eq.~(\ref{2channelSchr}). ``Short-range'' in our case means distances smaller than $1/\sqrt{t}$ (see below). For small $g_2$ one can neglect the logarithmic and imaginary terms in the denominator of Eq.~(\ref{Vertex}) arriving at $\Gamma\approx g_2-2\pi r_*|{\bf k}-{\bf k}'|$. The name coupling constant is thus attached to $g_2$ with the reservation that the neglected logarithm can become important for exponentially small energies $E\sim t\exp(-4\pi/|g_2|)$, in particular, when looking for poles of the scattering amplitude: as $g_2$ approaches zero from below, there is an exponentially weakly bound state with the binding energy $\varepsilon_0 = 4t\exp(4\pi/g_2)$.

Close to the crossing point $g_2$ is proportional to $t-t_c$ with a positive coefficient: for $t<t_c$ the two-body interaction is attractive (supports a bound state) and vice versa. To see how sensitive $g_2$ is to variations of $t$ for different $r_*$ in Fig.~\ref{fig:tVSrs} we show the region $-0.4<g_2<0.4$ enclosed by dotted lines. Making $g_2$ small at large $r_*$ requires a more subtle tuning because of a potential barrier which separates the scattering continuum from the bound state localized in this case at small $r$. The barrier is given by $V_{\uparrow\uparrow}(r)$ and $V_{\uparrow\downarrow}(r)$, which are both $\propto r_*$ and positive for $r>\sqrt{2}$. We should note that our results become qualitative for $t\sim 1$ ($r_*> 2$ or so) when the thickness of the layers comes into play. One then has to consider a full three-dimensional Hamiltonian instead of the idealized two-dimensional model (\ref{Ham}).

For $r_{*}\ll 1$ we solve the problem analytically by substituting for $V_{\sigma\sigma'}$ zero-range (ZR) pseudopotentials with proper low-energy scattering properties \cite{Shih}. In terms of Bessel functions the two-body wave function reads
\begin{equation}\label{linearsup}
\Psi=(|+\rangle)^2[J_0(qr)-if_{\rm zr}(q)H_0(qr)/4]+C(|-\rangle)^2K_0(\kappa r),
\end{equation}
where $q$ is the collision momentum and $\kappa=\sqrt{4t-q^2}$. The scattering amplitude
\begin{equation}\label{fzerorange}
f_{\rm zr}(q)=\frac{2\pi}{\ln\frac{\kappa}{q}-\ln\frac{\kappa^2}{E_{\uparrow\downarrow}}\ln\frac{\kappa^2}{E_{\uparrow\uparrow}}/\ln\frac{\kappa^4}{E_{\uparrow\downarrow}E_{\uparrow\uparrow}}+i\frac{\pi}{2}}
\end{equation} 
and coefficient $C=(2\pi)^{-1}f_{\rm zr}(q)\ln\frac{E_{\uparrow\downarrow}}{E_{\uparrow\uparrow}}/\ln\frac{\kappa^4}{E_{\uparrow\uparrow}E_{\uparrow\downarrow}}$ are determined from the ZR boundary conditions: $\phi_{\uparrow\uparrow}(r)\propto \ln(\sqrt{E_{\uparrow\uparrow}}r e^\gamma/2)$ and $\phi_{\uparrow\downarrow}(r)\propto \ln(\sqrt{E_{\uparrow\downarrow}}r e^\gamma/2)$, where $\gamma\approx 0.5772$ is the Euler constant, $E_{\uparrow\uparrow}=4\exp(-6\gamma)r_{*}^{-2}$ \cite{Ticknor}, and $E_{\uparrow\downarrow}\approx 4\exp(-8/r_{*}^2)$ \cite{Klawunn,Baranov,RemEb}. $f_{\rm zr}(0)$ vanishes for exponentially small $t=t_c=\sqrt{E_{\uparrow\downarrow}E_{\uparrow\uparrow}}/4\propto \exp(-4/r_{*}^2)$. We see that the ZR approximation is justified since typical length scales, $\sim 1/\sqrt{t_c}$, are exponentially large compared to the range of $V_{\sigma\sigma'}$. At $r_* \approx 0.7$ the ZR result for $t_c$ (dashed line in Fig.~\ref{fig:tVSrs}) deviates from the exact one by only about 10\%. The ZR approximation also predicts the dependence $g_2\approx 16\pi(\ln E_{\uparrow\uparrow}/E_{\uparrow\downarrow})^{-2}(t-t_c)/t_c$, valid for $(t-t_c)/t_c\ll 1$ and established by comparing Eq.~(\ref{fzerorange}) with the on-shell version of Eq.~(\ref{Vertex}). 

\begin{figure}
\centerline{\includegraphics[width=1\hsize,clip,angle=0]{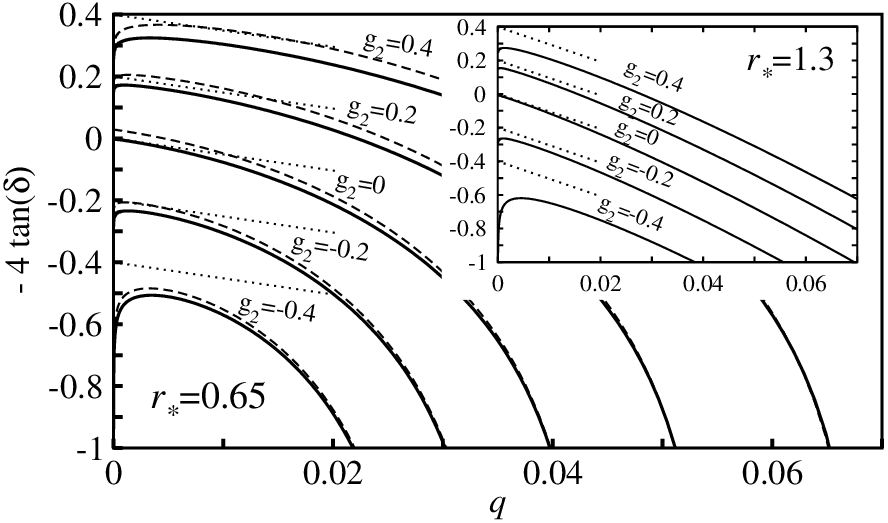}}
\caption{The function $-4\tan\delta$ vs. the collision momentum $q$ (solid lines) for $r_*=0.65$ for five values of $g_2$: $g_2=0.4$ ($t=2.4t_c$), $g_2=0.2$ ($t=1.5t_c$), $g_2=0$ ($t=t_c=5\times 10^{-4}$), $g_2=-0.2$ ($t=0.65t_c$), and $g_2=-0.4$ ($t=0.42t_c$). We also plot the approximations $-4\tan\delta\approx -4\tan\delta_{\rm zr}-8 r_* q$ (dashed lines) and $-4\tan\delta\approx g_2-8 r_* q$ (dotted lines). Inset shows the case $r_*=1.3$ for the same set of $g_2$, the tunneling amplitudes are $t=1.045t_c$, $t=1.023t_c$, $t=t_c=0.05$, $t=0.977t_c$, and $t=0.953t_c$. }
\label{fig:Tan}
\end{figure}

We find that for $r_*\lesssim 0.7$ and $q<2\sqrt{t}$ the exact $s$-wave scattering amplitude is well approximated by $f(q)\approx f_{\rm zr}(q)-8 r_* q$, where the last term is the $s$-wave component of $-2\pi r_*|{\bf q}-{\bf q}'|$ [see Eq.~(\ref{Vertex})] taken on the mass shell, $q=q'$. Figure~\ref{fig:Tan} shows $-4\tan\delta(q)$ (solid lines) and $-4\tan\delta_{\rm zr}(q)-8 r_* q$ (dashed lines) as functions of $q$ for $r_*=0.65$ and five values of $t$ close to $t_c$. Here we introduce the $s$-wave scattering phase shift $\delta$ and benefit from the relation $f=-4/(\cot\delta-i)\approx -4\tan\delta$ valid for small $\tan\delta$. We also present $f(q)\approx g_2-8 r_* q$ (dotted lines), which is the $s$-wave on-shell version of the approximation $\Gamma \approx g_2-2\pi r_*|{\bf q}-{\bf q}'|$. The inset in Fig.~\ref{fig:Tan} shows the case $r_*=1.3$ and we omit the ZR curves which are quite far off for this value of $r_*$.

\begin{figure}
\centerline{\includegraphics[width=0.9\hsize,clip,angle=0]{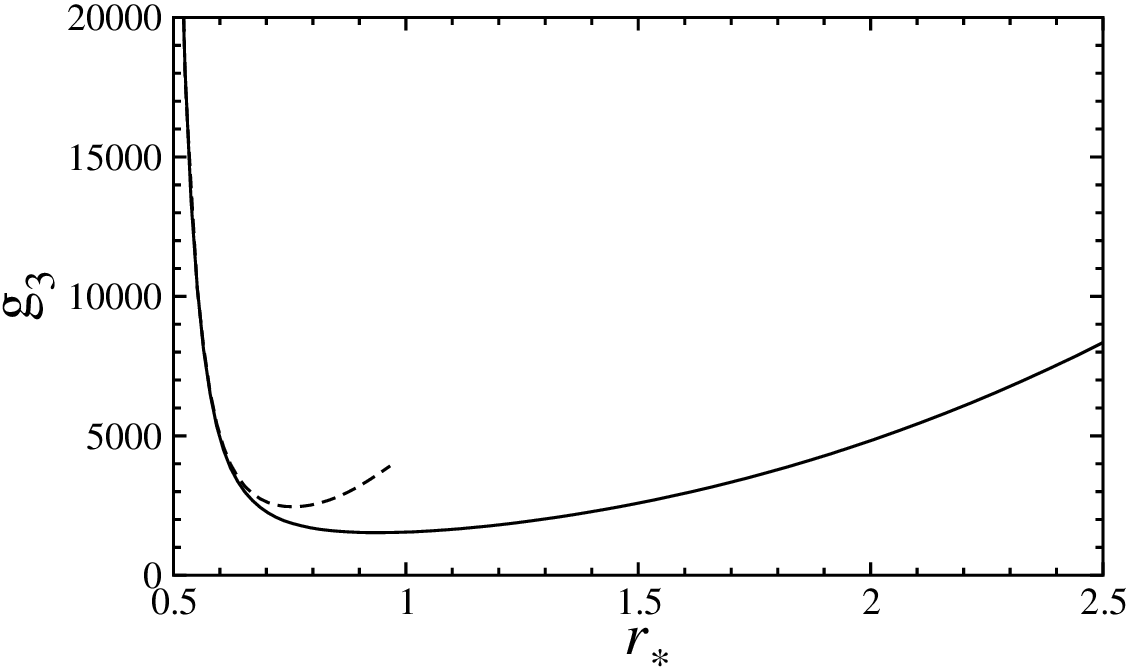}}
\caption{The three-body coupling constant $g_3$ (solid) versus $r_*$ for $t=t_c(r_*)$, i.e., for vanishing effective two-body interaction. The dashed line shows the result of the zero-range approximation (see text).}
\label{fig:Alpha}
\end{figure}

Let us now discuss the three-body problem. In three dimensions $g_3$ can be defined as the interaction energy shift of three condensed bosons in a unit volume with subtracted two-body contributions or, equivalently, in terms of the on-shell vertex functions at zero momenta,
\begin{equation}\label{alpha}
g_3=\langle free_{3} | \hat{V} | true_3 \rangle - 3 \langle free_2 | \hat{V} | true_2 \rangle,
\end{equation} 
where $\hat{V}$ is the interaction term in Eq.~(\ref{Ham}), $| free_n\rangle=(|+\rangle)^n$ is the ground state of $n$ noninteracting bosons, and $| true_n\rangle$ is the true zero energy $n$-body eigenstate of Eq.~(\ref{Ham}), normalized per unit volume. In two dimennsions, due to the two-dimensional kinematics \cite{RemLog}, the vertex functions should be considered at finite $E$. In our case the region where Eq.~(\ref{alpha}) remains approximately constant (thus defining $g_3$) is given by the inequalities $|g_2|\ll 1$ and $t\exp(-4 \pi/|g_2|)\ll E\ll t$. In Fig.~\ref{fig:Alpha} (solid line) we show $g_3$ calculated at the point $g_2=E=0$ which belongs to this region. We see that $g_3$ is always repulsive and rather large. It reaches $g_{3,{\rm min}}\approx 1530$ at $r_*=0.94$. In order to compute $| true_3\rangle$ we solve the three-body version of Eq.~(\ref{2channelSchr}) \cite{SM} by using the adiabatic hyperspherical approach \cite{Lin,Nielsen}.

The surprising enhancement of $g_3$ for small $r_*$ can be understood from the ZR analysis. In the case $q=0$ and $f_{\rm zr}(0)=0$ the $(|+\rangle)^2$-term in Eq.~(\ref{linearsup}) stands for two noninteracting bosons on a unit surface. The $(|-\rangle)^2$-term describes a ``bound'' pair of particles. Although $C=-2/\ln\sqrt{E_{\uparrow\uparrow}/E_{\uparrow\downarrow}}$ is small, the pair has an exponentially large size $\sim 1/\sqrt{t_c}$. The effective three-body force originates from the interaction of the third boson (in state $|+\rangle$) with either of these two particles, the collision energy being $\sim t_c$. Due to the bosonic symmetry the dominant $s$-wave interaction between $|+\rangle$- and $|-\rangle$-particles is given by the repulsive $V_{\uparrow\uparrow}$, which, at these energies, can be substituted by the corresponding vertex part $\approx 4\pi/\ln(E_{\uparrow\uparrow}/t_c)\approx 4\pi/\ln\sqrt{E_{\uparrow\uparrow}/E_{\uparrow\downarrow}}$. Overall, we obtain $g_3\propto t_c^{-1}(\ln\sqrt{E_{\uparrow\uparrow}/E_{\uparrow\downarrow}})^{-3}$, in which the factor $1/t_c$ provides the main (exponential) dependence on $r_*$. In a more rigorous perturbation theory \cite{SM} the two leading terms in the expansion of $g_3$ in the small parameter $\xi=1/\ln\sqrt{E_{\uparrow\uparrow}/E_{\uparrow\downarrow}}$ read $g_{3,\rm zr}=(24\pi^2/t_c)[\xi^3-3\ln(4/3)\xi^4+...]$ (dashed line in Fig.~\ref{fig:Alpha}).

That $g_3$ stays positive for all $r_*$ is a manifestation of the three-body frustration: at least two particles are on the same layer and, therefore, experience the repulsive potential $V_{\uparrow\uparrow}$. We claim that this mechanism protects the system against collapse and is responsible for a microscopic suppression of the local three-body correlation function, which clearly means suppressed three-body losses. It has been predicted \cite{Quemener,Micheli} and observed \cite{Jin} that inelastic processes can be suppressed by tightly confining dipoles to the two-dimensional geometry. Here we argue that the bilayer case with tunneling provides control over the interparticle interaction while preserving this suppression mechanism.

For RbK molecules with the interlayer distance $\lambda=532$nm the energy unit $\hbar^2/m\lambda^2$ corresponds to 280Hz or 13.4nK, which is rather small. However, for lighter molecules and shorter lattice periods \cite{Dulieu} one can gain an order of magnitude or more. It is also useful to keep in mind that similarly to spinor condensates the temperature requirement is rather loose \cite{RMPStamperKurnUeda}, more important is that the BEC chemical potential is smaller than $\hbar^2/m\lambda^2$, which is quite realistic.

The Hamiltonian (\ref{Ham}) in the rotating wave approximation also describes an atomic gas in which two hyperfine states ($\uparrow$ and $\downarrow$) are coupled by a resonant microwave or radio frequency field with the Rabi frequency $2t$. Here we can have significantly larger $t$ than in the bilayer dipolar case \cite{Zaccantiprivate}. However, in order to see three-body effects one has to control the three scattering lengths $a_{\uparrow\uparrow}$, $a_{\downarrow\downarrow}$, and $a_{\uparrow\downarrow}=a_{\downarrow\uparrow}$. The condition $a_{\uparrow\uparrow}=a_{\downarrow\downarrow}$ is not necessary, but for simplicity let us assume that it holds. 

An advantage of the ``atomic'' method is that the ZR approximation is essentially exact. In three dimensions we obtain \cite{SM}
\begin{equation}\label{f3D}
f^{(3D)}(q)=-\left[\frac{2-(a_{\uparrow\uparrow}+a_{\uparrow\downarrow})\kappa}{a_{\uparrow\uparrow}+a_{\uparrow\downarrow}-2a_{\uparrow\uparrow}a_{\uparrow\downarrow}\kappa}+iq\right]^{-1}.
\end{equation}
The effective scattering length vanishes for $\sqrt{t_c}=(1/a_{\uparrow\uparrow}+1/a_{\uparrow\downarrow})/4$. 
The three-body coupling constant can be calculated analytically for small $\xi=(a_{\uparrow\downarrow}+a_{\uparrow\uparrow})/(a_{\uparrow\downarrow}-a_{\uparrow\uparrow})$. Exactly at $t_c$ the leading term reads $g_3\approx 3\pi^2\xi^3/t_c^2 \approx 48\pi^2 a_{\uparrow\uparrow}^4/\xi$ \cite{SM}. This is by $1/\xi\gg 1$ larger than the usual three-body scaling $\propto a^4$ which appears only in the next order and, in particular, contains an imaginary part. Since it gives an estimate for the three-body recombination rate, the ratio of elastic to inelastic three-body interaction is $\propto 1/\xi$. Note, that in order to have a strong elastic three-body repulsion the scattering lengths are not required to be large. However, they should be very close in absolute values and have opposite signs: $a_{\uparrow\uparrow}>0$ and $a_{\uparrow\downarrow}=-a_{\uparrow\uparrow}(1+2\xi)$. Lysebo and Veseth \cite{Lysebo} predict rich possibilities for tuning interactions in between various hyperfine states of $^{39}$K. In particular, states $F=1,m_F=0$ and $F=1,m_F=-1$ in the magnetic field region from 50 to 60 Gauss can potentially give the desired effect, but one has to generalize the above theory to the case $a_{\uparrow\uparrow}\neq a_{\downarrow\downarrow}$.

In the two-dimensional case we can use the ZR formulas presented for the bilayer geometry with $E_{\sigma\sigma'}= B/(\pi l_0^2)\exp(\sqrt{2\pi}l_0/a_{\sigma\sigma'})$ \cite{IntCollPRA2001}, where $l_0$ is the transverse oscillator length and $B\approx 0.9$. Here the restriction on possible values of $a_{\sigma\sigma'}$ is softer than in three dimensions. It originates from the requirement to stay in the two-dimensional regime, $t_c\ll 1/l_0^2$, and reads $\exp[\sqrt{\pi/2}l_0(1/a_{\uparrow\uparrow}+1/a_{\uparrow\downarrow})]\ll 1$. 

The one-dimensional case can be realized by strongly confining atoms in two radial directions (or, for dipoles, in the bitube geometry). Introducing the one-dimensional scattering lengths $a_{1,\sigma\sigma'}=-l_0^2/a_{\sigma\sigma'}(1-{\cal C}'a_{\sigma\sigma'}/l_0)$, where ${\cal C}'\approx 1.0326$ \cite{Olshanii}, from Eq.~(\ref{2channelSchr}) we obtain \cite{SM}
\begin{equation}\label{f1D}
f^{(1D)}(q)=-\left[1+iq\frac{a_{1,\uparrow\uparrow}+a_{1,\uparrow\downarrow}-2a_{1,\uparrow\uparrow}a_{1,\uparrow\downarrow}\kappa}{2-(a_{1,\uparrow\uparrow}+a_{1,\uparrow\downarrow})\kappa}\right]^{-1}.\nonumber
\end{equation} 
The two-body coupling constant vanishes for $\sqrt{t_c}=1/(a_{1,\uparrow\uparrow}+a_{1,\uparrow\downarrow})$. At this point the three-body problem is equivalent to a two-dimensional one-body scattering by a potential with the range $\sim 1/\sqrt{t_c}$. The three-body interaction is conservative and repulsive for $a_{1,\uparrow\uparrow}<0$ reaching its maximum at $a_{1,\uparrow\uparrow}=0$ where the corresponding two-dimensional scattering length is of order $1/\sqrt{t_c}$. On the side $a_{1,\uparrow\uparrow}>0$ the interaction is stronger but is not conservative due to the recombination to dimer states of size $\sim a_{1,\uparrow\uparrow}$ \cite{SM}.

We now raise a few questions for future studies. Our intuition built on two-body interacting BECs may have to be reconsidered when three-body forces become important; fragmented condensation, rotonization, or (super)solidity of a three-body interacting gas with or without dipolar tails can not be {\it a priori} ruled out. Starting from the dilute droplet for $g_2<0$ and $g_3>0$ \cite{Bulgac} and further increasing the two-body attraction, will the system just shrink or eventually form a paired state? What is the interaction between pairs? In two dimensions, when $g_2$ is small there are two tetramers of exponentially large size \cite{Platter}, the local three-body interaction being a small perturbation. On the other hand, in the bilayer case with $r_*\gtrsim 1$ and $t=0$ the dimers are rather deeply bound and it is reasonable to assume that they repel each other. If and how the two tetramer states go into the dimer-dimer continuum with increasing $|g_2|$ remains an interesting few-body problem \cite{Volosniev}. Note that with the presented method we can also independently control the range of the three-body effective interaction, which can be important, in particular, for quantum Hall problems \cite{Jain}. Finally, the bilayer setup can be considered as a limiting case of a multilayer one in which the chemical potentials of two layers are significantly lowered. One can consider $N$ layers with generally different chemical potentials and thus try to engineer quite exotic effective few-body interactions (for example, $g_2=g_3=0$ and finite $g_4$). However, for $N\geq 3$ one should also be aware of inelastic three-body processes.

We thank N. Cooper, O. Dulieu, T. Jolicoeur, G. Shlyapnikov, and M. Zaccanti for fruitful discussions and acknowledge support by the IFRAF Institute.


\renewcommand{\theequation}{S\arabic{equation}}
\renewcommand{\thefigure}{S\arabic{figure}}

\setcounter{equation}{0}
\setcounter{figure}{0}
\vspace{1cm}
\centerline{\underline{\bf SUPPLEMENTAL MATERIAL}}

\section{Three-body Schr\"odinger equation}

In this section we derive the three-body Schr\"odinger equation, discuss its solution in the adiabatic hyperspherical approximation, and comment on the 2D kinematic problem and on the case of finite $g_2$ and $E$. 

The three sets of rescaled Jacobi coordinates ${\bf \Pi}_i=\{{\bf x}_i,{\bf y}_i\}$, $i=1,2,3$, are defined by
\begin{eqnarray}\label{JacobiCoord}
{\bf x}_i&=&(2{\boldsymbol{r}}_i-{\boldsymbol{r}}_j-{\boldsymbol{r}}_k)/\sqrt{3},\nonumber \\
{\bf y}_i&=&{\boldsymbol{r}}_k-{\boldsymbol{r}}_j,
\end{eqnarray}  
where $\{i,j,k\}$ are cyclic permutations of $\{1,2,3\}$. One can switch from one set to another by using the formulas
\begin{eqnarray}\label{Jacobi12}
{\bf x}_1&=&-{\bf x}_2/2+\sqrt{3}{\bf y}_2/2,\nonumber \\
{\bf y}_1&=&-\sqrt{3}{\bf x}_2/2-{\bf y}_2/2,\nonumber
\end{eqnarray}
and  
\begin{eqnarray}\label{Jacobi13}
{\bf x}_1&=&-{\bf x}_3/2-\sqrt{3}{\bf y}_3/2,\nonumber \\
{\bf y}_1&=&\sqrt{3}{\bf x}_3/2-{\bf y}_3/2.\nonumber
\end{eqnarray}

We look for the three-body wave function in the form
\begin{eqnarray}\label{Psi}
|true_3\rangle&=&(|\uparrow\rangle |\uparrow\rangle |\uparrow\rangle+|\downarrow\rangle |\downarrow\rangle |\downarrow\rangle)\phi_0({\bf \Pi}_1)\nonumber\\
&+&(|\uparrow\rangle |\downarrow\rangle |\downarrow\rangle + |\downarrow\rangle |\uparrow\rangle |\uparrow\rangle)\phi_1({\bf \Pi}_1)\nonumber\\
&+&(|\downarrow\rangle |\uparrow\rangle |\downarrow\rangle + |\uparrow\rangle |\downarrow\rangle |\uparrow\rangle)\phi_2({\bf \Pi}_1)\nonumber\\
&+& (|\downarrow\rangle |\downarrow\rangle |\uparrow\rangle +  |\uparrow\rangle |\uparrow\rangle |\downarrow\rangle)\phi_3({\bf \Pi}_1),\nonumber
\end{eqnarray}
where the bosonic symmetry requires 
\begin{equation}\label{Symmphi0}
\phi_0({\bf \Pi}_1)=\phi_0({\bf \Pi}_2)=\phi_0({\bf \Pi}_3)
\end{equation}
and
\begin{equation}\label{Symmphii}
\phi_i({\bf \Pi}_1)=\phi_1({\bf \Pi}_i),\; i=2,3.
\end{equation}
The Schr\"odinger equation for $\phi_0$ then reads
\begin{equation}\label{phi0}
[-\nabla^2_{{\bf \Pi}_1}+4t-E+\sum_{j=1}^3V_{\uparrow\uparrow}(y_j)]\phi_0-t\sum_{j=0}^3\phi_j=0
\end{equation}
and for $i\neq 0$
\begin{equation}\label{phii}
[-\nabla^2_{{\bf \Pi}_1}+4t-E+V_{\uparrow\uparrow}(y_i)+\sum_{j\neq i}V_{\uparrow\downarrow}(y_j)]\phi_i-t\sum_{j=0}^3\phi_j=0.
\end{equation}

For zero total angular momentum the configurational space of the problem is three-dimensional: the three-body wave function depends only on the hyperradius $\Pi=|{\bf \Pi}|$ and two (hyper)angles $\theta$ and $\phi$ defined by $x_1=\Pi\cos(\theta/2)$, $y_1=\Pi\sin(\theta/2)$, and $\phi$ is the angle between ${\bf x_1}$ and ${\bf y_1}$. Then the kinetic energy operator reads
\begin{eqnarray}\label{KineticEn}
-\nabla^2_{{\bf \Pi}_1}&=&-\frac{\partial^2}{\partial \Pi^2}-\frac{3}{\Pi}\frac{\partial}{\partial \Pi}\\\nonumber
&&-\frac{4}{\Pi^2}\left(\frac{\partial^2}{\partial\theta^2}+\cot\theta \frac{\partial}{\partial\theta}+\frac{1}{\sin^2\theta}\frac{\partial^2}{\partial \phi^2}\right).
\end{eqnarray}
Note that the hyperangular part is nothing else than the usual angular kinetic energy operator in 3D where $\theta$ and $\phi$ are polar and azimuthal angles, respectively. The fundamental hyperangular domain is 1/12 of the $\theta$-$\phi$ sphere defined by the inequalities $0<\phi<\pi/2$ and $\theta<\arcsin(1/\sqrt{1+\cos^2\phi})$. The boundary conditions for the wave function on the borders of the domain can be derived from Eqs.~(\ref{Symmphi0}-\ref{Symmphii}).

The set of four equations (\ref{phi0}-\ref{phii}) is solved numerically by using the adiabatic hyperspherical approach. Namely, we first diagonalize the hyperspherical part, i.e., we fix $\Pi$ and in the kinetic energy operator (\ref{KineticEn}) we keep only the second line. We choose a grid which has many points at small $\theta$ to allow for a better accuracy at large $\Pi$. Then, we solve the coupled channel hyperradial problem keeping typically 40 hyperspherical channels. However, we find that for our parameters 10 channels is sufficient for convergence.
 
The calculated $| true_3 \rangle$ is then substituted into
\begin{equation}\label{alphaSM}
g_3=\langle free_{3} | \hat{V} | true_3 \rangle - 3 \langle free_2 | \hat{V} | true_2 \rangle.
\end{equation} 
In principle, at $t=t_c$ the two-body contribution vanishes. However, let us mention an important technical detail. The brackets $\langle ... \rangle$ in the three-body term in Eq.~(\ref{alphaSM}) imply integration over the whole three-body configurational space. In practice, we deal with integrals of the type $\int_{\Pi<\Pi_{\rm max}}d^4 \Pi$ which, in fact, do not converge at $\Pi_{\rm max}\rightarrow \infty$ since the missing contribution from large hyperradii remains important for any $\Pi_{\rm max}$ because of the $1/y_i^3$ tails of the dipole-dipole potentials. The problem can, in principle, be cured analytically. However, a simpler solution is to treat the three- and two-body terms in Eq.~(\ref{alphaSM}) on equal footing, i.e., when calculating the two-body contribution we use the same region of integration $\Pi<\Pi_{\rm max}$ by introducing a third dummy particle, not interacting with the first two. In fact, as an additional consistency check we solve this three-body problem (with a dummy particle) by using the same machinery developed for the three-body case although with different symmetry requirements for the wave function. We can then double check the accuracy of the three-body numerical scheme since the two-body wave function is known very precisely from an ordinary ``two-body'' calculation.  

Finally, let us discuss in more detail the case $t\neq t_c$, i.e., finite $g_2$. For the sake of simplicity and readability we use the language of coupling constants, i.e., coefficients in front of powers of density in the low-density expansion of the energy functional. In 3D these coupling constants are the zero energy limits of the corresponding well-behaving vertex functions. Unfortunately, the formal zero energy limits of low-dimensional vertex functions are not physically reasonable. A significant part of the main text is sacrificed for sorting out this issue in the 2D case (the 1D case is discussed below in this Supplemental Material). For any finite two-body interaction the two-body wave function vanishes at finite interparticle distances in the limit $E\rightarrow 0$ because of the normalization condition. Therefore, strictly speaking, in the limit $E\rightarrow 0$ three interacting particles never approach each other to distances where they can feel the three-body potential. However, if $E$ is larger than $\propto \exp(-4 \pi/|g_2|)$, the normalization is no longer a problem, the particles do approach each other, and the three-body vertex function saturates to a finite value and then only weakly depends on $E$. We calculate the three-body wave function numerically for various values of $t$ and $E$ and see no dramatic changes in its shape as one crosses the point $g_2=0$ (apart from expected changes related to the finite collision momentum). The results are consistent with the fact that the three-body effective interaction acts at distances (hyperradii) $\sim 1/\sqrt{t}$ whereas the system feels the two-body interaction only at distances $\propto \exp(2 \pi/|g_2|)$. Therefore, one can explicitly introduce $g_3$ as a ``constant'' in the region defined by the inequalities $|g_2|\ll 1$ and $t\exp(-4 \pi/|g_2|)\ll E\ll t$. Note that this region includes the point $g_2=E=0$.

\section{Three-body problem in the zero-range approximation. 2D case}

In this section we derive the Skorniakov and Ter-Martirosian (STM) equations for the 2D four-channel three-body problem in the zero-range approximation (see Ref.~\cite{PetrovLesHouches} for a general introduction to this method). We then solve the STM equations and obtain the three-body coupling constant $g_{3,\rm zr}$ by using a systematic expansion with respect to the small parameter $\xi=1/\ln\sqrt{E_{\uparrow\uparrow}/E_{\uparrow\downarrow}}$.  

In the zero-range approximation the interaction potentials $V_{\uparrow\uparrow}$ and $V_{\uparrow\downarrow}$ in Eqs.~(\ref{phi0}-\ref{phii}) are substituted by boundary conditions on the wave functions $\phi_i$. Namely, we have for $i=1,2,3$ 
\begin{align}
&\phi_0|_{y_i\rightarrow 0}\propto  \ln(\sqrt{E_{\uparrow\uparrow}}y_i e^\gamma/2),\label{Boundary1}\\
&\phi_i|_{y_i\rightarrow 0}\propto  \ln(\sqrt{E_{\uparrow\uparrow}}y_i e^\gamma/2),\label{Boundary2}\\
&\phi_i|_{y_{j\neq i}\rightarrow 0}\propto \ln(\sqrt{E_{\uparrow\downarrow}}y_j e^\gamma/2).\label{Boundary3}
\end{align}
The proportionality symbols are understood in the sense that, for example, Eq.~(\ref{Boundary1}) fixes the ratio $A_2/A_1=\ln(\sqrt{E_{\uparrow\uparrow}} e^\gamma/2)$ in the small-$y_1$ expansion $\phi_0\approx A_1\ln(y_1)+A_2$.

By introducing a new set of wave functions $\{\psi_i\}$
\begin{equation}\label{psi}
\left(\begin{array}{l}\psi_0\\
\psi_1\\\psi_2\\\psi_3\end{array}\right)=\left(\begin{array}{rrrr}1&1&1&1\\
-1&1&0&0\\
-1&0&1&0\\
-1&0&0&1\end{array}\right)
\left(\begin{array}{l}\phi_0\\
\phi_1\\\phi_2\\\phi_3\end{array}\right)
\end{equation}
we write Eqs.~(\ref{phi0}-\ref{phii}) in the form
\begin{equation}\label{psi0}
(-\nabla^2_{{\bf \Pi}_1}-E)\psi_0=\sum_{j=1}^3f_0({\bf x}_j)\delta({\bf y}_j)
\end{equation}
and, for $i\neq 0$,
\begin{equation}\label{psii}
(-\nabla^2_{{\bf \Pi}_1}+4t-E)\psi_i=f_1({\bf x}_i)\delta({\bf y}_i)+\sum_{j\neq i}f_2({\bf x}_j)\delta({\bf y}_j).
\end{equation}
The functions $f_0$, $f_1$, and $f_2$ are defined by 
\begin{equation}\label{f}
f_i({\bf x}_1)=-2\pi \lim_{y_1\rightarrow 0}\left[y_1\frac{\partial\psi_i({\bf x}_1,{\bf y}_1)}{\partial y_1}\right],\; i=0,1,2.
\end{equation}
One can check that the right hand sides in Eqs.~(\ref{psi0}-\ref{psii}) together with the definition (\ref{f}) allow for the appearance of log-type singularities of $\psi$ at small $y_i$ consistent with the symmetry conditions (\ref{Symmphi0}-\ref{Symmphii}) (the sets $\{\psi_i\}$ and $\{\phi_i\}$ satisfy the same symmetry requirements).

We now invert Eqs.~(\ref{psi0}-\ref{psii}) and express $\psi_i$ as functionals of $f_i$:
\begin{widetext}
\begin{equation}\label{psi0throughf}
\psi_0({\bf \Pi}_1)=\psi_{00}({\bf \Pi}_1)+\sum_{j=1}^3\int G_E[\sqrt{({\bf x}_j-{\bf x})^2+y_j^2}]f_0({\bf x})d^2x,
\end{equation}
\begin{equation}\label{psiithroughf}
\psi_i({\bf \Pi}_1)=\int G_{E-4t}[\sqrt{({\bf x}_i-{\bf x})^2+y_i^2}]f_1({\bf x})d^2x+\sum_{j\neq i}\int G_{E-4t}[\sqrt{({\bf x}_j-{\bf x})^2+y_j^2}]f_2({\bf x})d^2x,
\end{equation}
\end{widetext}
where $\psi_{00}$ is a general (properly symmetrized) solution of the homogeneous Eq.~(\ref{psi0}). It plays the role of the incoming (free) wave. The Green function $G_E$ is the solution of $(-\nabla^2_{{\bf \Pi}}-E)G_E({\bf \Pi})=\delta({\bf \Pi})$ corresponding to the outgoing wave (for $E>0$). We do not have free terms in Eq.~(\ref{psiithroughf}) since we assume $E<4t$. In fact, in the four-dimensional ${\bf \Pi}$-space the wave functions $\psi_{i\neq 0}$ are localized in the areas defined by $|y_j|\lesssim 1/\sqrt{4t-E}$, i.e., they correspond to the virtually excited ``bound'' pairs (see discussion in the main text). Therefore, the outgoing scattered wave is given solely by the sum in the right hand side of Eq.~(\ref{psi0throughf}). 

In the case $E=0$ and $t=t_c=\sqrt{E_{\uparrow\uparrow}E_{\uparrow\downarrow}}/4$ the two-body scattering is absent, the three-body problem is equivalent to the 4D scattering by a finite-range potential, and $\psi_{00}({\bf \Pi}_1)\equiv \psi_{00}={\rm const}$. The wave function $\psi_0$ at large hyperradii reads 
\begin{equation}\label{psi0energy}
\psi_0 = \psi_{00}+3G_0(\Pi_1)\int f_0({\bf x})d^2x=\psi_{00}+3\frac{\int f_0({\bf x})d^2x}{4\pi^2 \Pi_1^2}
\end{equation}
and the three-body coupling constant equals
\begin{equation}\label{alphazrres}
g_3 = -(9/4)\int f_0({\bf x})d^2x/\psi_{00}.
\end{equation}
Equation~(\ref{alphazrres}) can be derived from Eq.~(\ref{psi0energy}) by switching from the potential to kinetic energy operators in the definition $g_3=\langle free_{3} | \hat{V} | true_3 \rangle$ and then using the Gauss-Ostrogradsky theorem which allows one to express $g_3$ in the form of an integral of $\nabla_{{\bf \Pi}_1}\psi_0$ over a four-dimensional sphere with infinitely large radius where we substitute the asymptote (\ref{psi0energy}). One should also take into account an additional factor $3/4$ which comes from the Jacobian determinant of the coordinate transformation (\ref{JacobiCoord}). Alternatively, one can simply note that the asymptote (\ref{psi0energy}) is the zero energy eigenstate of the four-dimensional problem of scattering by the pseudopotential $U_3$ defined by
\begin{equation}\label{3bodyPseud}
U_3\psi=g_3\delta(\sqrt{3}{\bf x}_1/2)\delta({\bf y}_1){\rm Reg}\psi,
\end{equation}
where ${\rm Reg}\psi=\partial[\Pi_1^2\psi({\bf \Pi}_1)]/\partial\Pi_1^2|_{\Pi_1\rightarrow 0}$.

The function $f_0$ is calculated in the following manner. We first obtain the functions $\phi_i$ by inverting Eq.~(\ref{psi}),
\begin{equation}\label{phi}
\left(\begin{array}{l}\phi_0\\\phi_1\\\phi_2\\\phi_3\end{array}\right)=\frac{1}{4}\left(\begin{array}{rrrr}1&-1&-1&-1\\
1&3&-1&-1\\
1&-1&3&-1\\
1&-1&-1&3\end{array}\right)
\left(\begin{array}{l}\psi_0\\\psi_1\\\psi_2\\\psi_3\end{array}\right),
\end{equation}
and substituting Eqs.~(\ref{psi0throughf}-\ref{psiithroughf}) into the right hand side of Eq.~(\ref{phi}). Coupled integral equations for $f_i$ (STM equations) are then derived by applying the boundary conditions (\ref{Boundary1}-\ref{Boundary3}). As we explain in the main text and in the end of the previous section of this Supplemental Material, if $t\neq t_c$, one has to work with finite energies in order to avoid the 2D kinematic issue. Here for simplicity we present the final equations for the particular case $t=t_c=\sqrt{E_{\uparrow\uparrow}E_{\uparrow\downarrow}}/4$ where we can safely set $E=0$. It is also convenient to switch from $f_i({\bf x})$ to the Fourier transformed functions $\tilde f_i({\bf p})=\int f_i({\bf x})\exp(-i\sqrt{4t_c}{\bf p}{\bf x})d^2x$, where we measure the momentum in units of $\sqrt{4t_c}$. Then the STM equations read
\begin{align}
&\tilde{f}_0= \xi [\hat{L}_{-1}\tilde{f}_1-\ln(1+p^2)(2\tilde{f}_2-\tilde{f}_1)],\label{STM1}\\
&\tilde{f}_1 =  \xi [-\hat{L}_{-1}\tilde{f}_2+\ln(1+p^2)\tilde{f}_1],\label{STM2}\\
&2\tilde{f}_2-\tilde{f}_1 =  \xi (\hat{L}_{0}-2\ln p)\tilde{f}_0+\xi(4\pi^3\psi_{00}/t_c)\delta({\bf p}),\label{STM3}
\end{align}
where the integral operator $\hat{L}_\epsilon$ is defined by
\begin{equation}
\hat{L}_\epsilon \tilde f({\bf p})=\frac{2}{\pi} \int \frac{\tilde{f}({\bf k})d^2 k}{p^2+k^2+{\bf pk}-3\epsilon/4}
\end{equation}
and $\xi = 1/\ln\sqrt{E_{\uparrow\uparrow}/E_{\uparrow\downarrow}}$. In the dipolar case this parameter is small as required for the validity of the zero-range approximation. Accordingly, we have arranged different terms in Eqs.~(\ref{STM1}-\ref{STM3}) for the following iterative procedure. Starting from the zeroth order $\tilde{f}_0^{(0)}=\tilde{f}_1^{(0)}=\tilde{f}_2^{(0)}=0$ the next order approximation is obtained by substituting the previous one into the right hand side of Eqs.~(\ref{STM1}-\ref{STM3}). In particular, the first order approximation is $\tilde{f}_0^{(1)}=\tilde{f}_1^{(1)}=0$, $\tilde{f}_2^{(1)}= \xi(2\pi^3\psi_{00}/t_c)\delta({\bf p})$, the second -- $\tilde{f}_0^{(2)}=0$, $\tilde{f}_1^{(2)}=2\tilde{f}_2^{(2)}=-\xi^2(4\pi^2\psi_{00}/t_c)/(p^2+3/4)$, etc. By continuing this procedure up to the fourth iteration we obtain
\begin{equation}\label{f_0}
\tilde{f}_0(0)=-(32\pi^2\xi^3\psi_{00}/3t_c)(1-3\ln(4/3)\xi+...),
\end{equation}
from which we get $g_{3,\rm zr}$ claimed in the paper.

Note that in contrast to the initial bilayer problem, the zero-range version of the purely repulsive potential $V_{\uparrow\uparrow}$ supports a bound state. In fact, this leads to two two-body bound states in our multi-channel model. One of them is in the spinor channel $(|\uparrow\rangle |\uparrow\rangle-|\downarrow\rangle |\downarrow\rangle)$ which, on the two-body level, is decoupled from $(|+\rangle)^2$ and $(|-\rangle)^2$ and has been completely neglected because of the $\uparrow$-$\downarrow$ symmetry of the Hamiltonian. Its binding energy is exactly $E_{\uparrow\uparrow}$. The second one corresponds to the pole of the zero-range scattering amplitude
\begin{equation}\label{fzerorangeSM}
f^{(2D)}(q)=\frac{2\pi}{\ln\frac{\kappa}{q}-\ln\frac{\kappa^2}{E_{\uparrow\downarrow}}\ln\frac{\kappa^2}{E_{\uparrow\uparrow}}/\ln\frac{\kappa^4}{E_{\uparrow\downarrow}E_{\uparrow\uparrow}}+i\frac{\pi}{2}}.
\end{equation} 
Its energy satisfies the relation
\begin{equation}\label{epsilon}
\ln(\epsilon/\sqrt{E_{\uparrow\downarrow}E_{\uparrow\uparrow}})\ln(1+\epsilon/\sqrt{E_{\uparrow\downarrow}E_{\uparrow\uparrow}})=(\ln\sqrt{E_{\uparrow\uparrow}/E_{\uparrow\downarrow}})^2,
\end{equation}
which for small $\xi$ also gives $\epsilon\approx E_{\uparrow\uparrow}$. Formally, in the zero-range approximation both of these states can be populated by three-body recombination. The phenomenon is characterized by the appearance of poles of functions $f_i$ at momenta $p_0\approx \exp(1/2\xi)$ and, consequently, by an imaginary contribution to $\tilde{f}_0(0)$ and $g_3$. One can show that these imaginary parts are $\propto \exp(-1/2\xi)$ and, accordingly, never appear in the Taylor expansion (\ref{f_0}) \cite{RemAsymptoticExp}. In the ``atomic'' case these two states can also be spurious. For example, for positive and small $a_{\uparrow\uparrow}\ll l_0$ the actual two-body bound states are three-dimensional and their energies have nothing to do with the quantity $E_{\uparrow\uparrow}= B/(\pi l_0^2)\exp(\sqrt{2\pi}l_0/a_{\uparrow\uparrow})$, which is just a characteristics of the 2D two-body scattering at low collision energy. In this case the three-body recombination has the usual 3D scaling $\propto a_{\uparrow\uparrow}^4$ and is small.

\section{3D case}

In this section we generalize our zero-range theory to the 3D case. 

The two-body wave function in the zero-range approximation reads
\begin{equation}\label{linearsup3D}
\Psi=(|+\rangle)^2\left[\frac{\sin(qr)}{qr}+f^{(3D)}(q)\frac{e^{iqr}}{r}\right]+C(|-\rangle)^2\frac{e^{-\kappa r}}{r}.
\end{equation}
Consequently the functions $\phi_{\uparrow\uparrow}$ and $\phi_{\uparrow\downarrow}$ equal
\begin{equation}\label{phi3D}
\phi_{\uparrow\uparrow/\uparrow\downarrow}(r)=\frac{\sin(qr)}{qr}+f^{(3D)}(q)\frac{e^{iqr}}{r}\pm C \frac{e^{-\kappa r}}{r}.
\end{equation}
The scattering amplitude
\begin{equation}\label{f3DSM}
f^{(3D)}(q)=-\left[\frac{2-(a_{\uparrow\uparrow}+a_{\uparrow\downarrow})\kappa}{a_{\uparrow\uparrow}+a_{\uparrow\downarrow}-2a_{\uparrow\uparrow}a_{\uparrow\downarrow}\kappa}+iq\right]^{-1}
\end{equation}
and coefficient
\begin{equation}\label{C3DSM}
C=\frac{a_{\uparrow\downarrow}-a_{\uparrow\uparrow}}{2-(a_{\uparrow\uparrow}+a_{\uparrow\downarrow})\kappa+iq(a_{\uparrow\uparrow}+a_{\uparrow\downarrow}-2a_{\uparrow\uparrow}a_{\uparrow\downarrow}\kappa)}
\end{equation}
are obtained by applying the zero-range boundary conditions
\begin{equation}\label{BethePeierls3D}
\phi_{\sigma\sigma'}|_{r\rightarrow 0}\propto 1-a_{\sigma\sigma'}/r
\end{equation}
to Eq.~(\ref{phi3D}).

The three-body approach of the previous sections of this Supplemental Material is modified as follows. The coordinates ${\bf r}$, ${\bf x}$, and ${\bf y}$ are now three-dimensional and ${\bf \Pi}$ is six-dimensional. The dimension of the integrals should be changed accordingly. The boundary conditions (\ref{Boundary1}-\ref{Boundary2}) in the 3D case become
\begin{align}
&\phi_0|_{y_i\rightarrow 0}\propto  1-a_{\uparrow\uparrow}/y_i,\label{Boundary13D}\\
&\phi_i|_{y_i\rightarrow 0}\propto  1-a_{\uparrow\uparrow}/y_i,\label{Boundary23D}\\
&\phi_i|_{y_{j\neq i}\rightarrow 0}\propto  1-a_{\uparrow\downarrow}/y_j,\label{Boundary33D}
\end{align}
and the functions $f_i$ are defined by
\begin{equation}\label{fSTM3D}
f_i({\bf x}_1)=4\pi \lim_{y_1\rightarrow 0}y_1\psi_i({\bf x}_1,{\bf y}_1),\; i=0,1,2.
\end{equation}
An important simplification of the 3D case compared to the low-D ones is that Eq.~(\ref{alphaSM}) is meaningful in the limit $E\rightarrow 0$ and the three-body interaction is clearly distinguished from the two-body one even when the latter is finite. In 3D we have
\begin{equation}\label{alphazrres3D}
g_3 = -(9\sqrt{3}/8)\int [f_0({\bf x})-\lim_{{\bf x} \rightarrow \infty}f_0({\bf x})]d^3x/\psi_{00},
\end{equation}
where the subtracted part proportional to $\lim_{{\bf x} \rightarrow \infty}f_0({\bf x})$ is the (tripled) two-body contribution.

The STM equations are derived in the same manner as in the 2D case. Now we write them for $E=0$ and for arbitrary $t$ not necessarily equal to $t_c$, but we still measure momentum in units of $\sqrt{4 t_c}=(1/a_{\uparrow\uparrow}+1/a_{\uparrow\downarrow})/2$, 
\begin{align}
&\tilde{f}_0= \xi [\hat{L}_{-t/t_c}\tilde{f}_1+(1-\sqrt{t/t_c+p^2})(2\tilde{f}_2-\tilde{f}_1)],\label{STM13D}\\
&\tilde{f}_1  =  \xi [-\hat{L}_{-t/t_c}\tilde{f}_2-(1-\sqrt{t/t_c+p^2})\tilde{f}_1],\label{STM23D}\\
&2\tilde{f}_2-\tilde{f}_1  =  \xi (\hat{L}_{0}+1-p)\tilde{f}_0+\xi(2\pi^4\psi_{00}/t_c^2)\delta({\bf p}),\label{STM33D}
\end{align}
where $\xi=(a_{\uparrow\downarrow}+a_{\uparrow\uparrow})/(a_{\uparrow\downarrow}-a_{\uparrow\uparrow})$ and
\begin{equation}
\hat{L}_\epsilon \tilde f({\bf p})=\frac{2}{\pi^2\sqrt{3}} \int \frac{\tilde{f}({\bf k})d^3 k}{p^2+k^2+{\bf pk}-3\epsilon/4}.
\end{equation}
For small $\xi$ we can perform the same iterative procedure as in the 2D case up to the third iteration arriving at the leading order term
\begin{equation}\label{g33D}
g_3\approx 3\pi^2\xi^3/\sqrt{t_c^3t},
\end{equation}
which for $t=t_c$ gives the expression presented in the main text. In contrast to the 2D case we can not continue the expansion further because momenta of order $p\sim 1/\xi$ become important starting from the next iteration. This renders the problem (\ref{STM13D}-\ref{STM33D}) non-perturbative: the terms proportional to $\xi^4/t_c^2\propto a^4$ depend on the short-range Efimov physics, three-body parameters in the $\uparrow\uparrow\uparrow$ and $\uparrow\uparrow\downarrow$ three-body channels, etc. In particular, the imaginary part of $g_3$ is of order $\xi^4/t_c^2\propto a^4$ and is related to the three-body recombination to weakly and deeply bound dimer states. 

Finally, we note that by tuning $a_{\uparrow\downarrow}$, $a_{\uparrow\uparrow}$ and $t$ one can have {\it any} effective scattering length
\begin{equation}\label{aeff3DSM}
a_{\rm eff}=\frac{a_{\uparrow\uparrow}+a_{\uparrow\downarrow}-2a_{\uparrow\uparrow}a_{\uparrow\downarrow}\sqrt{4t}}{2-(a_{\uparrow\uparrow}+a_{\uparrow\downarrow})\sqrt{4t}}.
\end{equation}
In particular, $a_{\rm eff}=\infty$ is obtained for $\sqrt{t}=1/(a_{\uparrow\downarrow}+a_{\uparrow\uparrow})$. An interesting question then concerns the effective three-body parameter which governs the Efimov physics of the unitary Bose gas obtained in this manner. Since it depends on a number of parameters ($a_{\uparrow\downarrow}$, $a_{\uparrow\uparrow}$, and the three-body parameters in the $\uparrow\uparrow\uparrow$ and $\uparrow\uparrow\downarrow$ channels), it is likely that one can, in principle, control it. However, more important is whether one can significantly reduce its imaginary part, i.e., minimize the three-body relaxation rate. Unfortunately, we can not give an affirmative answer. We have tried to set $a_{\uparrow\uparrow}$ to zero, thus eliminating also the three-body parameter in the $\uparrow\uparrow\uparrow$ channel. Then $a_{\uparrow\downarrow}=1/\sqrt{t}$ sets the length scale: on longer length scales the system behaves as a usual spinless unitary Bose gas with the scaling factor $\approx 22.7$, on shorter length scales we deal with the Efimov physics in the $\uparrow\uparrow\downarrow$ channel, which is characterized by a quite different scaling factor $\approx 1986.1$. Clearly, by changing $a_{\uparrow\downarrow}$ (and $\sqrt{t}$ accordingly) we can change the effective three-body parameter. However, we find that the effective inelasticity parameter can not be much reduced compared to the one for the $\uparrow\uparrow\downarrow$ channel.

\section{1D case}

Let us now discuss the 1D case in more detail. In the zero-range approximation the scattering happens only in the even ($s$-wave) channel. The two-body wave function reads
\begin{equation}\label{linearsup1D}
\Psi=(|+\rangle)^2[\cos(qr)+f^{(1D)}(q)e^{iq|r|}]+C(|-\rangle)^2 e^{-\kappa |r|}
\end{equation}
and the functions $\phi_{\uparrow\uparrow}$ and $\phi_{\uparrow\downarrow}$ equal
\begin{equation}\label{phi1D}
\phi_{\uparrow\uparrow/\uparrow\downarrow}(r)=\cos(qr)+f^{(1D)}(q)e^{iq|r|}\pm C e^{-\kappa |r|}.
\end{equation}
The scattering amplitude
\begin{equation}\label{f1DSM}
f^{(1D)}(q)=-\left[1+iq\frac{a_{1,\uparrow\uparrow}+a_{1,\uparrow\downarrow}-2a_{1,\uparrow\uparrow}a_{1,\uparrow\downarrow}\kappa}{2-(a_{1,\uparrow\uparrow}+a_{1,\uparrow\downarrow})\kappa}\right]^{-1}
\end{equation} 
and coefficient
\begin{equation}\label{C1DSM}
C=\frac{iq(a_{\uparrow\uparrow}-a_{\uparrow\downarrow})}{2-(a_{\uparrow\uparrow}+a_{\uparrow\downarrow})\kappa+iq(a_{\uparrow\uparrow}+a_{\uparrow\downarrow}-2a_{\uparrow\uparrow}a_{\uparrow\downarrow}\kappa)}
\end{equation}
are obtained by applying the boundary conditions
\begin{equation}\label{BethePeierls1D}
\frac{\partial \phi_{\sigma\sigma'}}{\partial r}\Big|_{r\rightarrow 0^+}-\frac{\partial \phi_{\sigma\sigma'}}{\partial r}\Big|_{r\rightarrow 0^-}=-\frac{2}{a_{1,\sigma\sigma'}}\phi_{\sigma\sigma'}(0)
\end{equation}
to Eq.~(\ref{phi1D}). The scattering amplitude (\ref{f1DSM}) corresponds to the two-body scattering by the pseudopotential $g_2(q) \delta(r)$, which depends on the scattering momentum $q$,
\begin{equation}\label{g31DSM}
g_2(q)=-2\frac{2-(a_{1,\uparrow\uparrow}+a_{1,\uparrow\downarrow})\sqrt{4t-q^2}}{a_{1,\uparrow\uparrow}+a_{1,\uparrow\downarrow}-2a_{1,\uparrow\uparrow}a_{1,\uparrow\downarrow}\sqrt{4t-q^2}}.
\end{equation}
The approximation $g_2(q)=g_2(0)=g_2$ can be made for $q\ll \sqrt{t}$ if $(a_{1,\uparrow\uparrow}- a_{1,\uparrow\downarrow})/(a_{1,\uparrow\uparrow}+a_{1,\uparrow\downarrow})$ is not anomalously small. The zero crossing occurs at $t=t_c$, where $\sqrt{t_c}=1/(a_{1,\uparrow\uparrow}+a_{1,\uparrow\downarrow})$. It is useful to note that at this point a finite momentum can not render the two-body interaction strong since the ratio $g_2(q)|_{t=t_c}/q$, which measures the ``strength'' of the interaction in 1D, is linear in $q$ and, therefore, remains small for small finite $q$.

The configurational space ${\bf \Pi}=\{x,y\}$ of the 1D three-body problem is two-dimensional. The effective three-body potential acts at distances $\Pi\lesssim 1/\kappa\sim 1/\sqrt{t}$ and originates from the interaction of the $(|-\rangle)^2$-component of the wave function (\ref{linearsup1D}) with the third particle in state $|+\rangle$. The effect of this finite range three-body interaction depends on the value of $g_2$ and the energy. For example, if $g_2=0$, the three-body scattering is equivalent to the 2D $s$-wave scattering by a finite range potential. The corresponding vertex function should scale as \cite{RemThreeBodyreduced} 
\begin{equation}\label{Vertex3body1}
\Gamma_3(E)\approx 2\sqrt{3}\pi/[\ln(\epsilon_3/E)+i\pi]
\end{equation}
for small $E$. When $g_2=\infty$ (fermionized or Tonks gas), the scattering is equivalent to the 2D $g$-wave scattering (angular momentum $l=3$) because the three-body wave function has six nodes on a circle drawn around the origin in the two-dimensional ${\bf \Pi}$-space. In this case the threshold law for the scattering amplitude is $\propto E^3$. Clearly, for finite $g_2$ the kinematic regime of scattering is determined by the ratio $g_2/\sqrt{E}$. For {\it any} finite $g_2$ the system will fermionize at energies $E\lesssim g_2^2$. Nevertheless, we can neglect two-body interactions for $\Pi<1/|g_2|$ and, if this radius is larger than the range of the three-body interaction $1/\sqrt{t}$, the quantity $\epsilon_3$ is meaningful. It is a continuous function of $t-t_c$ close to the zero crossing and can be used, for example, to construct a local three-body pseudopotential. Moreover, it characterizes the actual three-body interaction as opposed to the kinematic effects that happen at distances $\Pi \sim 1/|g_2|,1/\sqrt{E}$ where the structure of the three-body wave function changes solely due to the two-body interactions.

We thus calculate $\epsilon_3$ in the case $g_2=0$. If one insists on introducing the three-body coupling constant $g_3$, this can be done in the same manner as we do in the main text when defining $g_2$ in the 2D case. Namely,
\begin{equation}\label{Vertex3body2}
\Gamma_3(E)\approx 2\sqrt{3}\pi/[\ln(4t/E)+2\sqrt{3}\pi/g_3+i\pi],
\end{equation}
where $g_3$ is defined as $g_3=2\sqrt{3}\pi/\ln(\epsilon_3/4t)$ and is assumed to be small. Then in the first order Born approximation one can use the pseudopotential
\begin{equation}\label{3bodyPseud1D}
U_3=(2/\sqrt{3})g_3\delta({\bf \Pi})=g_3\delta(\sqrt{3}x/2)\delta(y).
\end{equation}
For example, the chemical potential of a weakly purely three-body interacting 1D (quasi)condensate can be estimated as $\mu\approx g_3n^2/2$, although it would be more appropriate to use the self-consistent equation $\mu \approx \Gamma_3(3\mu)n^2/2$ in analogy with the weakly two-body-interacting 2D Bose gas.

The 1D STM equations are derived in the same manner as in the 2D and 3D cases. The boundary conditions (\ref{Boundary1}-\ref{Boundary2}) in the 1D case read
\begin{align}
&\frac{\partial \phi_0}{\partial y_i}\Big|_{y_i\rightarrow 0^+}-\frac{\partial \phi_0}{\partial y_i}\Big|_{y_i\rightarrow 0^-}=-\frac{2}{a_{1,\uparrow\uparrow}}\phi_0|_{y_i=0},\label{Boundary11D}\\
&\frac{\partial \phi_i}{\partial y_i}\Big|_{y_i\rightarrow 0^+}-\frac{\partial \phi_i}{\partial y_i}\Big|_{y_i\rightarrow 0^-}=-\frac{2}{a_{1,\uparrow\uparrow}}\phi_i|_{y_i=0},\label{Boundary21D}\\
&\frac{\partial \phi_i}{\partial y_j}\Big|_{y_{j}\rightarrow 0^+}-\frac{\partial \phi_i}{\partial y_j}\Big|_{y_{j}\rightarrow 0^-}=-\frac{2}{a_{1,\uparrow\downarrow}}\phi_i|_{y_j=0},\;j\neq i,\label{Boundary31D}
\end{align}
and the functions $f_i$ are defined by
\begin{equation}\label{fSTM1D}
f_i({\bf x}_1)=-\frac{\partial \phi_i}{\partial y_1}\Big|_{y_1\rightarrow 0^+}+\frac{\partial \phi_i}{\partial y_1}\Big|_{y_1\rightarrow 0^-},\; i=0,1,2.
\end{equation}

In the low energy limit we can set $E=0$ in Eq.~(\ref{psiithroughf}). However, in Eq.~(\ref{psi0throughf}) we have to keep $E$ small but finite since the Green function in this case scales as $G_E(\Pi)\approx -\ln(\sqrt{-E} e^\gamma\Pi/2)/(2\pi)$. Then, (for $g_2=0$) the large-$\Pi$ asymptote of the three-body wave function reads
\begin{equation}\label{psi0energy1D}
\psi_0(\Pi) = \psi_{00}-\frac{3}{2\pi}\ln\frac{\sqrt{-E} e^\gamma\Pi}{2}\int f_0({\bf x})dx.
\end{equation}
On the other hand $\psi_0$ should be proportional to $\ln(\sqrt{\epsilon_3} e^\gamma\Pi/2)$ consistent with Eq.~(\ref{Vertex3body1}). We thus find
\begin{equation}\label{epsilon3}
\epsilon_3=-E\exp\left[-\frac{4\pi\psi_{00}}{3\int f_0({\bf x})dx}\right].
\end{equation}
We will now show that in the limit $E\rightarrow 0$ the dependence of $f_0$ on $E$ is such that $\epsilon_3$ is energy independent. 

The 1D STM equations read
\begin{align}
&\tilde{f}_0= \xi [\hat{L}_{-1}\tilde{f}_1+(1/\sqrt{1+p^2}-1)(2\tilde{f}_2-\tilde{f}_1)],\label{STM11D}\\
&\tilde{f}_1  =  \xi [-\hat{L}_{-1}\tilde{f}_2-(1/\sqrt{1+p^2}-1)\tilde{f}_1],\label{STM21D}\\
&2\tilde{f}_2-\tilde{f}_1  =  \xi (\hat{L}_{\epsilon}+1/\sqrt{p^2-\epsilon}-1)\tilde{f}_0+4\pi\xi\psi_{00}\delta(p),\label{STM31D}
\end{align}
where $\xi = (a_{1,\uparrow\uparrow}+ a_{1,\uparrow\downarrow})/(a_{1,\uparrow\downarrow}-a_{1,\uparrow\uparrow})$, $\tilde f_i(p)=\int f_i(x)\exp(-i\sqrt{4t_c}px)dx$, and we measure momenta in units of $\sqrt{4t_c}$. In Eq.~(\ref{STM31D}) $\epsilon =E/4t_c\ll 1$ and we have set $E$ to zero in Eqs.~(\ref{STM11D}) and (\ref{STM21D}). The integral operator in 1D is defined by
\begin{equation}\label{}
\hat{L}_\epsilon \tilde f(p)=\frac{\sqrt{3}}{\pi} \int \frac{\tilde{f}(k)d k}{p^2+k^2+pk-3\epsilon/4}.
\end{equation}
The $\epsilon$ dependence in Eqs.~(\ref{STM11D}-\ref{STM31D}) can be singled out by using the equations
\begin{equation}
\hat{L}_\epsilon 1 = 2/\sqrt{p^2-\epsilon}
\end{equation}
and
\begin{equation}
\lim_{\epsilon\rightarrow 0}\hat{L}_{-1} \frac{1}{\sqrt{p^2-\epsilon}} =F(p)-\frac{\sqrt{3}}{\pi}\frac{\ln(-\epsilon)}{p^2+3/4},
\end{equation}
where
\begin{equation}
F(p)=\frac{\sqrt{3}}{\pi}\frac{\ln(4p^2+3)}{p^2+3/4}+\frac{2p\;{\rm arccot}(\sqrt{3}\sqrt{p^2+1}/p)}{\pi(p^2+3/4)\sqrt{p^2+1}}.
\end{equation}
We then define a new function $\tilde h_2(p)$ by
\begin{equation}\label{f2throughh2}
\tilde f_2(p)=\tilde h_2(p)+\frac{\xi}{2}\left[\frac{3\tilde f_0(0)}{\sqrt{p^2-\epsilon}}+4\pi\psi_{00}\delta(p)\right],
\end{equation}
substitute Eq.~(\ref{f2throughh2}) into Eqs.~(\ref{STM11D}-\ref{STM31D}), and take the limit $\epsilon\rightarrow 0$ where it exists. The result is
\begin{widetext}
\begin{align}
&\tilde{f}_0(p)= \xi \{\hat{L}_{-1}\tilde{f}_1(p)+(1/\sqrt{1+p^2}-1)[2\tilde{h}_2(p)-\tilde{f}_1(p)]\}+3\xi^2(1/\sqrt{p^2+1}-1)|p|^{-1}\tilde f_0(0),\label{STM11Dreg}\\
&\tilde{f}_1(p)  =  \xi [-\hat{L}_{-1}\tilde{h}_2(p)-(1/\sqrt{1+p^2}-1)\tilde{f}_1(p)]-(3/2)\xi^2F(p)\tilde f_0(0)-2\sqrt{3}\xi^2\Lambda/(p^2+3/4),\label{STM21Dreg}\\
&2\tilde{h}_2(p)-\tilde{f}_1(p)  =  \xi (\hat{L}_{0}+1/|p|)[\tilde{f}_0(p)-\tilde f_0(0)]-\xi \tilde f_0(p),\label{STM31Dreg}
\end{align}
\end{widetext}
where $\Lambda=\psi_{00}-(3/4\pi)\tilde f_0(0)\ln(-\epsilon)$ and we can set it to 1. Then, Eqs.~(\ref{STM11Dreg}-\ref{STM31Dreg}) do not contain $\epsilon$, and, therefore, $\tilde f_0$, $\tilde f_1$, and $\tilde h_2$ do not depend on energy. The $\epsilon$ dependence is transfered to $\psi_{00}$ which now equals $\psi_{00}=1+(3/4\pi)\tilde f_0(0)\ln(-\epsilon)$. Equation~(\ref{psi0energy1D}) then becomes explicitly energy independent, $\psi_{0}(\Pi)\approx 1-(3/2\pi)\tilde f_0(0)\ln(\sqrt{4t_c}e^\gamma\Pi/2)$, and Eq.~(\ref{epsilon3}) reduces to $\epsilon_3=4t_c\exp[-4\pi/3\tilde f_0(0)]$.

Equations~(\ref{STM11Dreg}-\ref{STM31Dreg}) can be solved iteratively for small $\xi$ as we do for higher dimensions. In fact, in the purely 1D case there are no two-body bound states for $|\xi|<1$ and the three-body interaction is absolutely conservative (at least in the purely 1D model). The leading terms are
\begin{equation}\label{f_01D}
\tilde{f}_0(0)=-(8\xi^3/\sqrt{3})[1+(2/3-2\sqrt{3}/\pi)\xi+(3\sqrt{3}-5)\xi^2+...],
\end{equation}
which means that the three-body interaction is weakly repulsive (attractive) for small positive (negative) $\xi$. The interaction gradually increases with $|\xi|$ and $\epsilon_3$ becomes of order $t_c$ when $|\xi|\sim 1$. We have not looked at finite energies, but from the structure of the zero energy wave function it is clear that there are no trimer states for $\xi>0$ and the interaction can be considered as a soft core repulsion. In contrast, for $\xi<0$ there is a node of $\psi_0$ at large $\Pi$ and there is a trimer state, the binding energy of which for small $|\xi|$ equals $\epsilon_3$.   

The point $\xi=1$ on the repulsive side is equivalent to $a_{1,\uparrow\uparrow}=0$, i.e., particles of the same type are impenetrable. This gives a strong three-body repulsion. The repulsion can be increased even further by making $\xi$ somewhat larger than $1$ (small positive $a_{1,\uparrow\uparrow}$) and thus entering the so-called super-Tonks regime. However, similarly to the 2D and 3D cases, there are two two-body bound states of small size $\approx a_{1,\uparrow\uparrow}$, three-body recombination to which is possible. For $\xi<-1$ ($a_{1,\uparrow\downarrow}>0$), there is one such bound state. The rate of recombination to these states can be found numerically from Eq.~(\ref{STM11Dreg}-\ref{STM31Dreg}) but we leave this task for future studies.


\begin{thebibliography}{99}
\bibitem{RemSingleBoundState} We restrict ourselves to the case of a single bound state.
\bibitem{RemEb} In our calculations we use numerically exact $E_{\uparrow\downarrow}(r_*)$.
\bibitem{RemLog} The wave functions $| true_2\rangle$ and $| true_3\rangle$ vanish in the formal zero energy limit for any finite distance between two particles [check, for example, Eq.~(\ref{linearsup}) in the limit $q\rightarrow 0$]. Equation~(\ref{alpha}) then gives zero similarly to Eq.~(\ref{Vertex}) at $E=0$.
\end{thebibliography}
\begin{thebibliography}{99}
\bibitem{RemAsymptoticExp} In fact, for a given $\xi$ the iterative procedure diverges and the series (\ref{f_0}) is asymptotic, which is also consistent with the scaling of the correction $\propto \exp(-1/2\xi)$. We find that in the region of our interest ($r_*> 0.5$) the next two iterations do not improve the expansion (\ref{f_0}).
\bibitem{RemThreeBodyreduced} Equation~(\ref{Vertex3body1}) differs from the usual two-body 2D vertex function by the factor $\sqrt{3}/2$ originating from the Jacobian of the transformation (\ref{JacobiCoord}) or, equivalently, from the fact that the three-body reduced mass is $2/\sqrt{3}$ times larger than the two-body one.
\end{thebibliography}
\end{document}